\documentclass[12pt]{article}
\usepackage{epsfig,xspace}
\usepackage{cite}
\def\beq{\begin{equation}}
\def\eeq#1{\label{#1}\end{equation}}
\def\eeqn{\end{equation}}

\def\beqa{\begin{eqnarray}}
\def\eeqa#1{\label{#1}\end{eqnarray}}
\def\eeqan{\end{eqnarray}}

\let\bar=\overbar

\def\Dslash{\not{\hbox{\kern-4pt $D$}}}
\def\dslash{\not{\hbox{\kern-2pt $\del$}}}

\def\msb{{\bar{\ssstyle M \kern -1pt S}}}

\newcommand{\sNN}{$\sqrt{s_{NN}}$\xspace}
\newcommand{\sqrts}{$\sqrt{s}$\xspace}

\newcommand{\MeV}{MeV/$c$\xspace}
\newcommand{\GeV}{GeV/$c$\xspace}

\newcommand{\Fref}[1]{Figure~\ref{#1}}

\newcommand{\pp}{pp\xspace}

\newcommand{\Au}{Au+Au\xspace}
\newcommand{\Pb}{Pb+Pb\xspace}
\newcommand{\dAu}{$d$+Au\xspace}
\newcommand{\AplusA}{$A$+$A$\xspace}

\newcommand{\pT}{$p_{T}$\xspace}
\newcommand{\pikp}{$\pi^\pm$, K$^\pm$, and p($\bar{p}$)\xspace}

\def\Title#1{\begin{center} {\Large {\bf #1} } \end{center}}

\begin{document}

\Title{Results from ALICE}
\bigskip\bigskip


\begin{raggedright}  

{\it Christine Nattrass for the ALICE collaboration\index{Nattrass, C.}\\
Physics Department\\
University of Tennessee at Knoxville\\
Knoxville, TN 37996 USA }
\bigskip\bigskip
\end{raggedright}

\section{Abstract}

The ALICE experiment at the Large Hadron Collider at CERN is optimized to study the properties of the hot, dense matter created in high energy nuclear collisions in order to improve our understanding of the properties of nuclear matter under extreme conditions.  In 2009 the first proton beams were collided at the Large Hadron collider and since then data from proton-proton collisions at  \sqrts = 0.9, 2.36, 2.76, and 7 TeV have been taken.  Results from \pp collisions provide significant constraints on models.  In particular, results on strange particles indicate that Monte Carlo generators still have considerable difficulty describing strangeness production.  In 2010 the first lead nuclei were collided at \sNN = 2.76 TeV.    Results from \Pb demonstrate suppression of particle production relative to that observed in \pp collisions, consistent with expectations based on data available at lower energies.

\section{Introduction}


A Large Ion Collider Experiment (ALICE)~\cite{Kuijer:2002xq,Aamodt:2008zz} is a general purpose detector optimized to measure the bulk properties of the matter created in \Pb collisions.  The primary goal of studies of ultra-relativistic heavy-ion physics is the study of nuclear matter at extreme temperatures and energy densities.  At sufficiently high energy densities nuclear matter transitions from ordinary nuclear matter to a phase of deconfined quarks and gluons, called the Quark Gluon Plasma (QGP).  The creation of a QGP is possible in high energy nuclear collisions~\cite{Back:2004je,Adcox:2004mh,Arsene:2004fa,Adams:2005dq}.  However these collisions present experimental challenges due to the large track densities in central \Pb collisions.  It is therefore necessary for tracking detectors to have fine granularity.  In addition, many of the observables which can be used to determine the properties of the QGP are flavor and mass dependent.  It is therefore crucial to have information on particle identification over as wide a kinematic range as possible.  ALICE's capabilities for low momentum tracking and particle identification allow ALICE to perform measurements in \pp collisions complementary to those that other experiments at the LHC emphasize.  The LHC has provided data from \pp collisions at center of mass energies \sqrts = 0.9, 2.36, 2.76, and 7 TeV and \Pb collisions at a center of mass energy per nucleon of \sNN = 2.76 TeV.

\begin{figure}[htb]
\begin{center}
\epsfig{file=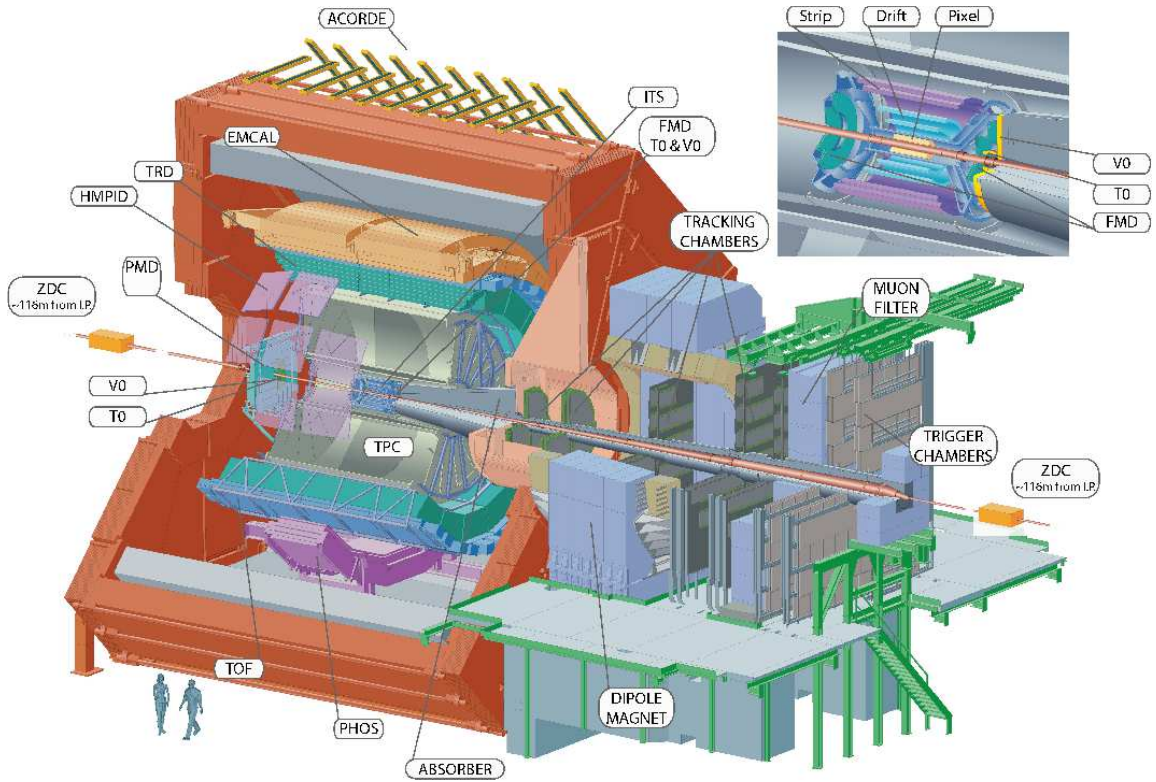,width=6in}
\caption{The ALICE detector comprises 18 detector systems.  Acronyms are defined in the text.}
\label{fig:aliceschematic}
\end{center}
\end{figure}

The ALICE detector, shown in \Fref{fig:aliceschematic}, is 16 m in diameter and 26 m long and weighs approximately 10,000 tons.  Since ALICE is designed for measurements of events with high track densities, ALICE has precision detectors but with limited acceptance, focusing on midrapidity ($|\eta|<$ 0.9).  The central detectors sit inside of the L3 magnet, which produces a nominal magnetic field of 0.5 T.  The Inner Tracking System (ITS) surrounds the beam pipe and consists of a Silicon Pixel Detector (SPD), Silicon Drift Detector (SDD), and Silicon Strip Detector (SSD).  These silicon detectors provide $<$ 100 $\mu$m resolution of tracks' distance of closest approach to the primary vertex for tracks with pseudorapidity $|\eta|<$ 0.9.  The ITS is capable of multiplicity measurements for tracks with transverse momentum \pT $>$ 0.050 \GeV and tracking for particles with \pT $>$ 0.100 \GeV.  A large Time Projection Chamber (TPC) surrounds the inner tracking system, extending from a radius of approximately 85 cm from the beam pipe to a radius of approximately 250 cm from the beam pipe.  It is approximately 5 m along the beam pipe, providing tracking for $|\eta|<$ 0.9 with momentum resolution $\Delta p_{T}/p_{T}<$ 1\% for tracks completely contained in the TPC acceptance.

Both the ITS and the TPC are capable of particle identification through energy loss.  The Time-Of-Flight (TOF) detector covers $|\eta|<$ 0.9 and can separate pions and kaons up to a momentum of 2.5 \GeV and protons and kaons up to a momentum of 4 \GeV.  The High Momentum Particle Idenfication Detector (HMPID) is a ring-imaging Cherenkov detector optimized to extend $\pi$/K discrimination to 3 \GeV and K/p discrimination up to 5 \GeV.  The HMPID covers $|\eta|<$ 0.6 in pseudorapidity and 1.2$^\circ<\phi<$ 58.8$^\circ$.  ALICE has two electromagnetic calorimeters, the Photon Spectrometer (PHOS) and the Electromagnetic Calorimeter (EMCAL).  PHOS is 4.6 m from the vertex, is made of scintillating crystals (PbWO$_4$) with high resolution and granularity, and is optimized to measure photons.  The EMCAL is a lead scintillator sampling calorimeter optimized for studies of jets.  PHOS covers $|\eta|<$ 0.12 and 100$^\circ$ in azimuth and the EMCAL covers $|\eta|<$ 0.7 and 107$^\circ$ in azimuth.  A COsmic Ray DEtector (ACORDE) provides triggers for cosmic ray events for calibration and alignment as well as studies of cosmic ray physics.  The Muon Spectrometer sits between 2$^\circ$ and 9$^\circ$ from the beam pipe and triggers on and tracks muons for studies of heavy quark particles, mainly through $J/\psi$ and $\Upsilon$.   In addition there are several smaller detectors at small angles for triggering and multiplicity measurements, the Zero Degree Calorimeter (ZDC), Photon Multiplicity Detector (PMD), Forward Meson Detector (FMD), T0, and V0.

\section{Results}
ALICE is able to measure the bulk properties of \pp and \Pb events accurately because of its capabilities for precision low momentum tracking and particle identification.  These capabilities allow precision measurements of particle multiplicities and transverse energy. Since separation of \pikp is possible over a wide kinematic region, the ALICE detector is complementary to those of other LHC experiments.  Measurements of strange particles which undergo weak decays (K$^0_S$, $\Lambda$, $\Xi$, $\Omega$), resonances (e.g., $\phi$, K$^*$), and charmed hadrons which decay hadronically (e.g., D$^\pm$, D$^0$, D$^{\pm}_S$) can be improved substantially since the combinatorial background can be reduced substantially by identifying the decay daughters.

\subsection{Bulk properties}

ALICE's low momentum tracking capabilities allow measurements of track multiplicities down to \pT = 50 \MeV, limiting the extrapolation necessary to measure charged particle multiplicities (d$N_{ch}$/d$\eta$).  The energy dependence of multiplicities in \pp collisions at \sqrts = 0.9~\cite{:2009dt,Aamodt:2010ft}, 2.36~\cite{Aamodt:2010ft}, and 7 TeV~\cite{Aamodt:2010pp} is described well by a power law in energy, $s^{0.1}$ and multiplicities are generally above model predictions.  Models underpredict the increase in multiplicity as a function of collision energy and most of the discrepancy with models is in the high multiplicity tails of the distribution of events.  These measurements have already been used to refine models for particle production in \pp collisions.

\begin{figure}[htb]
\begin{center}
\epsfig{file=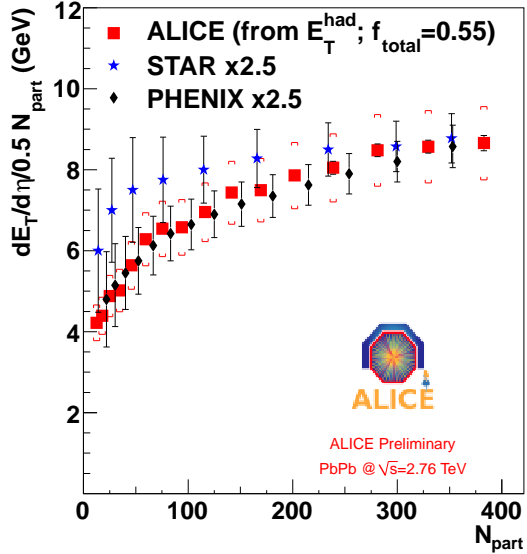,height=3in}
\caption{Transverse energy over the number of participating nucleons, $N_{\mathrm{part}}$ divided by two from ALICE~\cite{Loizides:2011ys} compared to PHENIX~\cite{Adcox:2001ry} and STAR~\cite{Adams:2004cb} scaled by a factor of 2.5 to compare the shape of the $N_{\mathrm{part}}$ dependence.  $f_{\mathrm{total}}$ is the correction factor to correct for neutral particles which were not measured.}
\label{fig:et}
\end{center}
\end{figure}

Multiplicity measurements in \Pb collisions substantially constrain models.  Predictions for multiplicities in central \Pb collisions at \sNN = 2.76 TeV made after data from \Au collisions at \sNN = 200 GeV at the Relativistic Heavy Ion Collider (RHIC) were available ranged from 1000-1700~\cite{Abreu:2007kv}.  ALICE data constrained this to 1584 $\pm$ 4 (stat.) $\pm$ 76 (syst.)~\cite{Aamodt:2010pb}, leading to an energy dependence of d$N_{ch}$/d$\eta$ described well by a power law in energy of $s^{0.15}$ in heavy-ion collisions.  Centrality dependent studies showed that the data at the LHC have the same shape as a function of the number of participating nucleons observed at RHIC~\cite{Aamodt:2010cz}, with the data within error up to an overall scaling factor.  Transverse energy measurements, shown in \Fref{fig:et}, demonstrate the same trend~\cite{Loizides:2011ys,Adcox:2001ry,Adams:2004cb}.  Charged particle multiplicity and transverse energy measurements indicate that the energy densities reached at the LHC are approximately three times larger than those produced at RHIC~\cite{Collaboration:2011rta}.

\subsection{Charged particles}
In \pp collisions measurements of charged particle spectra indicate that models which describe the particle multiplicity in \pp collisions well still struggle to describe the shape of the momentum spectrum~\cite{Aamodt:2010my}.  Models are particularly poor at describing low momentum (\pT$<$ 500 \MeV) particles.  Identified \pikp spectra show that models fail to describe the particle composition, systematically underestimating the production of both kaons and protons at high momenta~\cite{Aamodt:2011zj}.  The combination of ALICE's particle identification capabilities with the precision low momentum tracking allows proton identification and rejection of secondary protons.  This enables accurate measurements of the $\bar{p}$/p ratio in \pp collisions at both \sqrts = 0.9 (0.957 $\pm$ 0.006 $\pm$ 0.014) and 7 TeV (0.991 $\pm$ 0.005 $\pm$ 0.014).  This ratio is described by models well~\cite{Aamodt:2010dx}.

\begin{figure}[htb]
\begin{center}
\epsfig{file=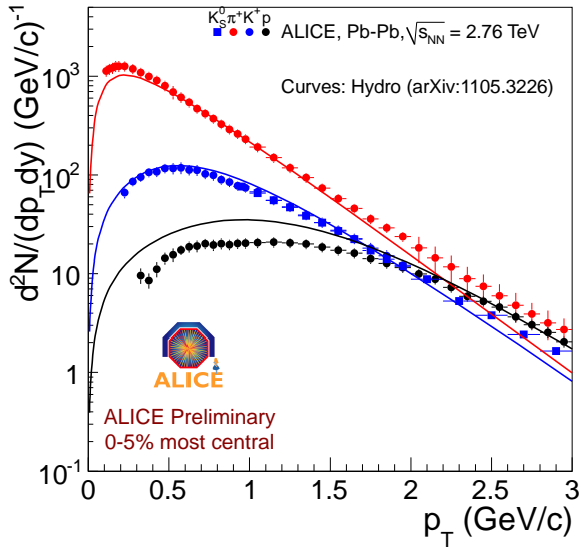,height=3in}
\caption{Spectra of identified particles compared to a hydrodynamical model for particle production~\cite{Floris:2011ru}}
\label{fig:pikp}
\end{center}
\end{figure}

Suppression of high momentum particle production is expected in \AplusA collisions due to interactions of hard partons with the hot, dense medium.  This suppression is often quantified by reporting the nuclear modification factor:
\begin{equation}
 R_{AA}(p_T) = \frac{  (1/N_{\mathrm{evt}}^{AA}) d^2N^{AA}/d\eta dp_{T}  }{ <N_{\mathrm{coll}}> (1/N_{\mathrm{evt}}^{pp})  d^2N^{pp}/d\eta dp_{T} },
\end{equation}
\noindent the ratio of the particle yield in \AplusA collisions ($1/N_{\mathrm{evt}}^{AA}) d^2N^{AA}/d\eta dp_{T}$) to that in \pp collisions ($(1/N_{\mathrm{evt}}^{pp})  d^2N^{pp}/d\eta dp_{T}$) at the same energy scaled by the number of binary collisions in the \AplusA collisions ($<N_{\mathrm{coll}}>$).  If \Pb collisions were just a superposition of nucleon-nucleon collisions, the nuclear modification factor would be one at high \pT, where particle production is dominated by hard parton scattering.  $R_{\mathrm{AA}}$ unidentified charged particles reaches values as low as ~0.15, lower than the suppression of ~0.2 observed at RHIC~\cite{Aamodt:2010jd}.  Early comparisons of identified \pikp spectra in heavy ion collisions shown in \Fref{fig:pikp} indicate that while models provide a reasonable description of pions and kaons, they fail to describe both the shape and the yield of protons and anti-protons~\cite{Floris:2011ru}.

%
\subsection{Strange particles}
Strange particles which decay weakly can be identified by the reconstruction of their decay vertices and identification of the decay daughters improves measurements by helping reduce the combinatorial background.  Models fail to describe either the shape or the yield of strange particles, underestimating kaon spectra by as much as a factor of two above 1 \GeV and $\Lambda$ and $\bar{\Lambda}$ spectra by as much as a factor of three above 1 \GeV in \pp collisions at \sqrts = 0.9 TeV~\cite{Aamodt:2011zz}.  Similar discrepancies between Monte Carlo event generators and kaon and $\Lambda$ ($\bar{\Lambda}$) spectra are observed at \sqrts = 7 TeV.  Production of $\Xi^{\pm}$ and $\Omega^{\pm}$ is underestimated by as much as a factor of four and ten, respectively~\cite{Chinellato:2011yn}.  However, models are considerably better at describing production of the $\phi$ resonance~\cite{Pulvirenti:2011xs}.  While Monte Carlo generators have substantial difficulties describing the particle yields of strange particles, at RHIC energies statistical models for particle production were able to describe the ratios of particle yields in \pp collisions well.  Instead, a comparison of the ALICE data to the THERMUS model indicate that at the LHC the model is currently unable to describe the data~\cite{Floris:2011ru}.

\begin{figure}[htb]
\begin{center}
\epsfig{file=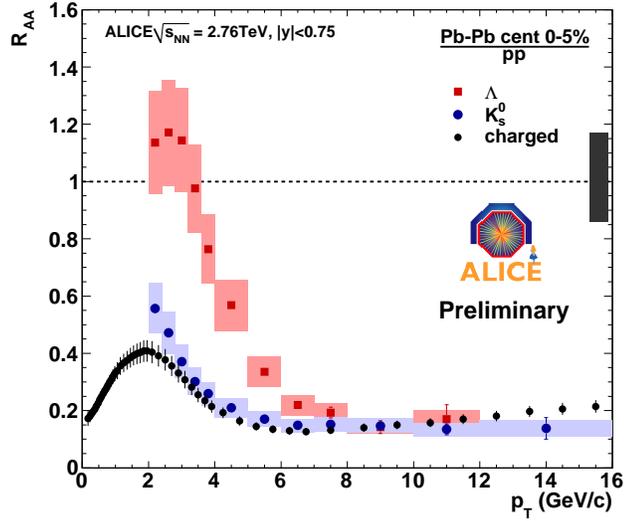,height=3in}
\caption{Nuclear modification factor ($R_{\mathrm{AA}}$) of $\Lambda$ and $K^0_S$ compared to unidentified hadrons as a function of \pT~\cite{Collaboration:2011xk}}
\label{fig:strange}
\end{center}
\end{figure}

For strange particles in central \Pb collisions, the nuclear modification factor of $\Lambda$ and $K^0_S$ compared to unidentified hadrons shown in \Fref{fig:strange} approach the same value at high \pT, indicating that mass effects are less significant at higher momenta~\cite{Collaboration:2011xk}.  
Furthermore, the ratio of $\Lambda$/$K^0_S$ exhibits an enhancement over that observed in \pp collisions.  This enhancement is comparable to that observed in central \Au collisions at \sNN = 200 GeV, although about 20\% larger~\cite{Collaboration:2011xk}. 


\subsection{Heavy flavors}
\begin{figure}[htb]
\begin{center}
\epsfig{file=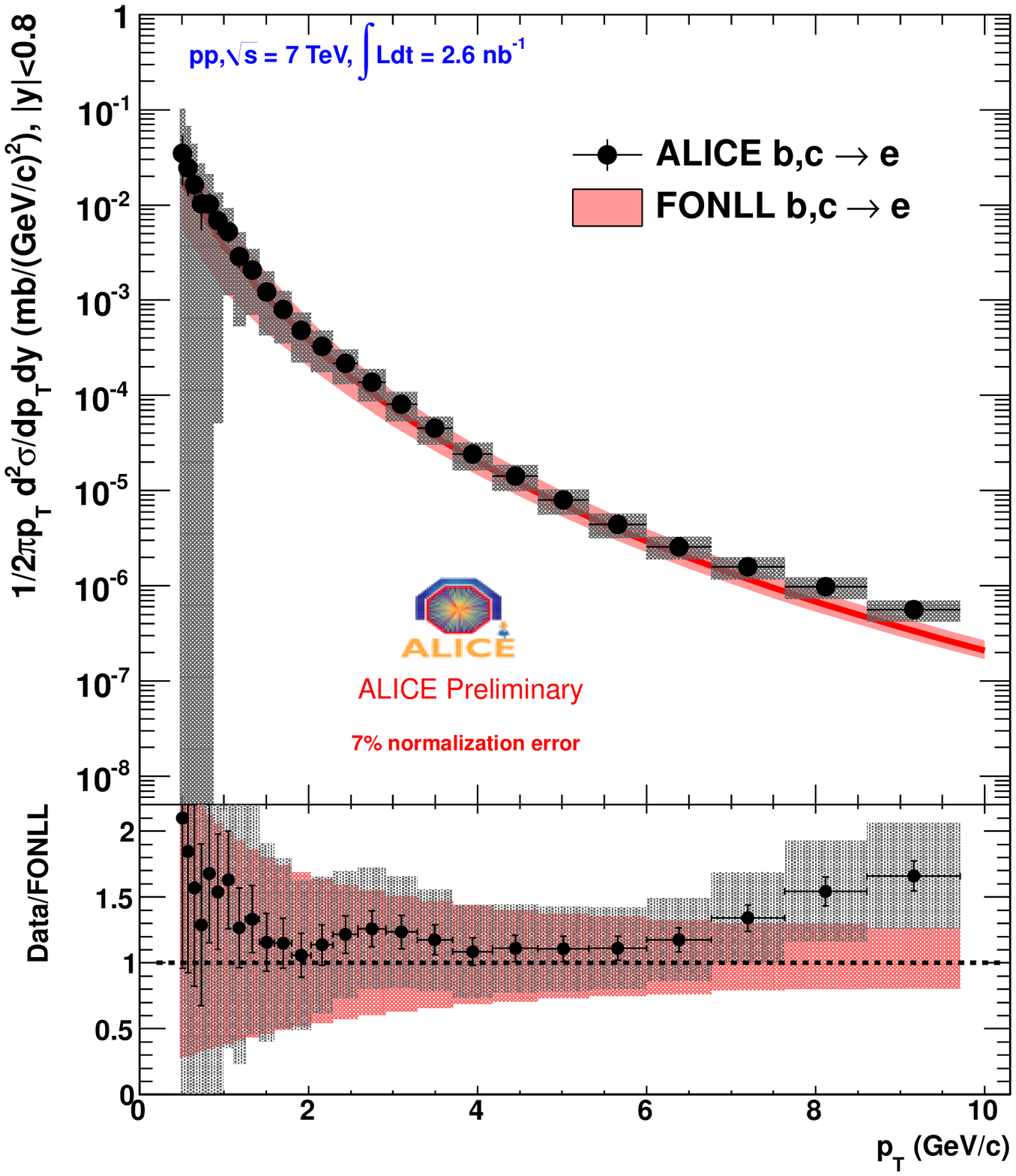,height=5in}
\caption{Spectrum of electrons from semi-leptonic decays of heavy quarks compared to Fixed Order Next-to-Leading-Log perturbative QCD calculations~\cite{Masciocchi}}
\label{fig:charm}
\end{center}
\end{figure}

ALICE allows multiple measurements of heavy flavor production.  Hadrons containing heavy quarks have substantial cross sections for semi-leptonic decays and ALICE can measure heavy quark production various decay channels.  Studies of non-photonic electrons in \pp collisions at \sqrts = 7 TeV are consistent with Fixed Order Next-to-Leading-Log perturbative QCD calculations~\cite{Masciocchi}, as shown in \Fref{fig:charm}.  Heavy flavor can also be measured through reconstruction of heavy quarks which decay hadronically (e.g., D$^{\pm}$, D$^0$) and through quarkonia which decay into dileptons (e.g., $J/\psi$).  Studies of the nuclear modification factor for D$^{\pm}$ and D$^0$ mesons indicate that the suppression of heavy flavor in \Pb collisions at \sNN = 2.76 TeV is comparable to the suppression of light flavors, similar to what was observed at RHIC~\cite{Dainese:2011vb}.  The nuclear modification factor of the $J/\psi$ in \Pb collisions at \sNN = 2.76 is higher than that of the D$^{\pm}$ and D$^0$ mesons~\cite{Dainese:2011vb}, however, the interpretation of this result is complicated by the observation of the suppression of the $J/\psi$ in \dAu collisions~\cite{Adare:2010fn}.



\section{Conclusions}
Since the first collisions in November 2009, the LHC has delivered \pp collisions at \sqrts = 0.9, 2.36, 2.76, and 7 TeV and \Pb collisions at \sNN = 2.76 TeV.  The ALICE detector is able to make precise measurements of particle multiplicities, transverse energy, identified particle spectra, strange particle spectra, and heavy flavor.  Measurements in both \pp and \Pb collisions have already constrained models considerably, restricting multiplicities and indicating that Monte Carlo generators need substantial refinement to be able to describe strange particle production.


\begin{thebibliography}{99}

\bibitem{Kuijer:2002xq}
  P.~Kuijer [ Alice Collaboration ],

\bibitem{Aamodt:2008zz}
  K.~Aamodt {\it et al.} [ ALICE Collaboration ],
  JINST {\bf 3}, S08002 (2008).

\bibitem{Back:2004je}
  B.~B.~Back, M.~D.~Baker, M.~Ballintijn, D.~S.~Barton, B.~Becker, R.~R.~Betts, A.~A.~Bickley, R.~Bindel {\it et al.},
  Nucl.\ Phys.\  {\bf A757}, 28-101 (2005).
  [nucl-ex/0410022].
\bibitem{Adcox:2004mh}
  K.~Adcox {\it et al.} [ PHENIX Collaboration ],
  Nucl.\ Phys.\  {\bf A757}, 184-283 (2005).
  [nucl-ex/0410003].
\bibitem{Arsene:2004fa}
  I.~Arsene {\it et al.} [ BRAHMS Collaboration ],
  Nucl.\ Phys.\  {\bf A757}, 1-27 (2005).
  [nucl-ex/0410020].
\bibitem{Adams:2005dq}
  J.~Adams {\it et al.} [ STAR Collaboration ],
  Nucl.\ Phys.\  {\bf A757}, 102-183 (2005).
  [nucl-ex/0501009].


\bibitem{:2009dt}
  K.~Aamodt {\it et al.}  [ALICE Collaboration],
  Eur.\ Phys.\ J.\  C {\bf 65}, 111 (2010)
  [arXiv:0911.5430 [hep-ex]].

\bibitem{Aamodt:2010ft}
  K.~Aamodt {\it et al.} [ ALICE Collaboration ],
  Eur.\ Phys.\ J.\  {\bf C68}, 89-108 (2010).
  [arXiv:1004.3034 [hep-ex]].

\bibitem{Aamodt:2010pp}
  KAamodt {\it et al.} [ ALICE Collaboration ],
  Eur.\ Phys.\ J.\  {\bf C68}, 345-354 (2010).
  [arXiv:1004.3514 [hep-ex]].

\bibitem{Abreu:2007kv}
  N.~Armesto, (ed.), N.~Borghini, (ed.), S.~Jeon, (ed.), U.~A.~Wiedemann, (ed.), S.~Abreu, V.~Akkelin, J.~Alam, J.~L.~Albacete {\it et al.},
  J.\ Phys.\ G {\bf G35}, 054001 (2008).
  [arXiv:0711.0974 [hep-ph]].

\bibitem{Aamodt:2010pb}
  BAbelev {\it et al.} [ The ALICE Collaboration ],
  Phys.\ Rev.\ Lett.\  {\bf 105}, 252301 (2010).
  [arXiv:1011.3916 [nucl-ex]].

\bibitem{Aamodt:2010cz}
  K.~Aamodt {\it et al.} [ ALICE Collaboration ],
  Phys.\ Rev.\ Lett.\  {\bf 106}, 032301 (2011).
  [arXiv:1012.1657 [nucl-ex]].

\bibitem{Loizides:2011ys}
  C.~Loizides, f.~t.~A.~collaboration,
    [arXiv:1106.6324 [nucl-ex]].

\bibitem{Adcox:2001ry}
  K.~Adcox {\it et al.}  [PHENIX Collaboration],
  Phys.\ Rev.\ Lett.\  {\bf 87}, 052301 (2001)
  [arXiv:nucl-ex/0104015].

\bibitem{Adams:2004cb}
  J.~Adams {\it et al.} [ STAR Collaboration ],
  Phys.\ Rev.\  {\bf C70}, 054907 (2004).
  [nucl-ex/0407003].

\bibitem{Collaboration:2011rta}
  A.~T.~f.~Collaboration,
  arXiv:1107.1973 [nucl-ex].


\bibitem{Aamodt:2010my}
  KAamodt {\it et al.} [ ALICE Collaboration ],
  Phys.\ Lett.\  {\bf B693}, 53-68 (2010).
  [arXiv:1007.0719 [hep-ex]].


\bibitem{Aamodt:2011zj}
  K.~Aamodt {\it et al.} [ ALICE Collaboration ],
  Eur.\ Phys.\ J.\  {\bf C71}, 1655 (2011).
  [arXiv:1101.4110 [hep-ex]].

\bibitem{Aamodt:2010dx}
  A.~:: K.~Aamodt {\it et al.} [ ALICE Collaboration ],
  Phys.\ Rev.\ Lett.\  {\bf 105}, 072002 (2010).
  [arXiv:1006.5432 [hep-ex]].


\bibitem{Aamodt:2010jd}
  K.~Aamodt {\it et al.} [ ALICE Collaboration ],
  Phys.\ Lett.\  {\bf B696}, 30-39 (2011).
  [arXiv:1012.1004 [nucl-ex]].

\bibitem{Floris:2011ru}
  M.~Floris and f.~t.~A.~Collaboration,
  arXiv:1108.3257 [hep-ex].



\bibitem{Aamodt:2011zz}
  K.~Aamodt, A.~Abrahantes Quintana, D.~Adamova, A.~M.~Adare, M.~M.~Aggarwal, G.~Aglieri Rinella, A.~G.~Agocs, S.~Aguilar Salazar {\it et al.},
  Eur.\ Phys.\ J.\  {\bf C71}, 1594 (2011).
  [arXiv:1012.3257 [hep-ex]].

\bibitem{Chinellato:2011yn}
  D.~D.~Chinellato, f.~t.~A.~Collaboration,
  [arXiv:1106.6314 [hep-ex]].

\bibitem{Pulvirenti:2011xs}
  A.~Pulvirenti, f.~t.~A.~Collaboration,
  [arXiv:1106.4230 [hep-ex]].


\bibitem{Collaboration:2011xk}
  I.~B.~f.~t.~A.~Collaboration,
    [arXiv:1109.4807 [hep-ex]].

\bibitem{Masciocchi}
  S.~Masciocchi, f.~t.~A.~Collaboration,
    [arXiv:1109.6436 [nucl-ex]].


\bibitem{Dainese:2011vb}
  A.~Dainese and f.~t.~A.~Collaboration,
  arXiv:1106.4042 [nucl-ex].

\bibitem{Adare:2010fn}
  A.~Adare, S.~Afanasiev, C.~Aidala, N.~N.~Ajitanand, Y.~Akiba, H.~Al-Bataineh, J.~Alexander, A.~Angerami {\it et al.},
  
  [arXiv:1010.1246 [Unknown]].



%
%







\end{thebibliography}
\end{document}